# Asymmetric stress engineering of dense dislocations in brittle superconductors for strong vortex pinning

Meng Han[1,13], Chiheng Dong[1,2,13], Chao Yao[1,2], Zhihao Zhang[3], Qinghua Zhang[4], Yue Gong[5], He Huang[1], Dongliang Gong[1], Dongliang Wang[1,2], Xianping Zhang[1,2], Fang Liu[6], Yuping Sun[7], Zengwei Zhu[8], Jianqi Li[4], Junyi Luo[9], Satoshi Awaji[9], Xiaolin Wang[10], Jianxin Xie[3,*], Hideo Hosono[11,12,*], and Yanwei Ma[1,2,*]

[1] Key Laboratory of Applied Superconductivity, Institute of Electrical Engineering, Chinese Academy of Sciences, Beijing 100190, People's Republic of China

[2] University of Chinese Academy of Sciences, Beijing 100049, People's Republic of China

[3] Institute for advanced materials and technology, University of Science and Technology Beijing, Beijing, 100083, People's Republic of China

[4] Beijing National Laboratory for Condensed Matter Physics, Institute of Physics, Chinese Academy of Sciences, Beijing 100190, People's Republic of China

[5] CAS Key Laboratory of Standardization and Measurement for Nanotechnology, CAS Center for Excellence in Nanoscience, National Center for Nanoscience and Technology, Beijing 100190, People's Republic of China

[6] Institute of Plasma Physics, Chinese Academy of Sciences, Hefei 230031, People's Republic of China

[7] High Magnetic Field Laboratory, Chinese Academy of Sciences, Hefei 230031, People's Republic of China

[8] Wuhan National High Magnetic Field Center, Huazhong University of Science and Technology, Wuhan 430074, China.

[9] Institute for Materials Research, Tohoku University, Sendai 980-8577, Japan

[10] Institute for Superconducting & Electronic Materials, University of Wollongong, North Wollongong, New South Wales 2500, Australia

[11] MDX Research Center for Element Strategy, Mailbox S2-13, Tokyo Institute of Technology, 4259 Nagatsuta-cho, Midori-ku, Yokohama 226-8503, Japan.

[12] MANA Center, National Institute for Materials Science, 1-1 Namiki, Tsukuba, Ibaraki 305-0044, Japan

[13] These authors contributed equally: Meng Han, Chiheng Dong

* Corresponding author. E-mail: jxxie@mater.ustb.edu.cn; hosono@mces.titech.ac.jp; ywma@mail.iee.ac.cn

**Abstract:** Large lossless currents in high-temperature superconductors (HTS) critically rely on dense defects with suitable size and dimensionality to pin vortices, with dislocations being particularly effective due to their 1D geometry to interact extensively with vortex lines. However, in non-metallic compounds such as HTS with rigid lattices, conventional deformation methods typically lead to catastrophic fracture rather than dislocation-mediated plasticity, making it a persistent challenge to introduce dislocations at high density. Here, an asymmetric stress field strategy is proposed using extrusion to directly nucleate a high-density of dislocations in HTS by activating shear-driven lattice slip and twisting under superimposed hydrostatic compression. As demonstrated in iron-based superconductors (IBS), atomic displacements of nearly one angstrom trigger the formation of tilted dislocation lines with a density approaching that of metals. With further structural refinement, these dislocations serve as strong pinning centers that lead to a fivefold enhancement in the current-carrying capacity of IBS at 33 T, along with low anisotropy and a large irreversibility field. This work not only establishes a scalable route to engineer pinning landscapes in HTS, but also offers a generalizable framework for manipulating dislocation structures in rigid crystalline systems.

# 1. Introduction

The development of advanced technologies such as compact fusion reactors, high-field magnetic resonance imaging (MRI) systems, and next-generation particle accelerators critically relies on superconductors capable of sustaining large critical current densities ($J_c$) under high magnetic fields[1–3]. In practical superconductors, such large supercurrents are only achievable when quantized magnetic vortices are effectively immobilized by defects that act as pinning centers[4]. Theoretically, the key principle of vortex pinning is to disrupt the long-range coherence of the superconducting state in regions with lateral sizes close to the coherence length ($\xi$), which is generally 1-3 nm in high-temperature superconductors (HTS)[5]. Consequently, effective pinning requires nanoscale defects with appropriate dimensionality to match vortex lines[6]. Among defects with varieties of morphologies, dislocations stand out as intrinsic, lattice-compatible pinning centers. With lateral sizes comparable to vortex cores and one-dimensional geometry, they enable extended and coherent vortex-defect interactions, providing strong pinning forces that boost $J_c$.

Dislocations are commonly found in metals and alloys due to the presence of multiple slip systems, high atomic mobility, and non-directional metallic bonding, all of which facilitate dislocation nucleation and motion[7]. In contrast, the strong ionic and covalent bonds in HTS form a rigid crystal lattice that is inherently less tolerant of high strain energy. The directional nature of these bonds severely limits the mobility and rearrangement of atoms[8]. Conventional deformation methods make it difficult to construct a stress state with high hydrostatic pressure superimposed with sufficient shear stress. As a result, they tend to cause bond breakage and brittle fracture, rather than activating dislocation-mediated plasticity. Figure 1 compares the response of crystal lattices, modeled here as Rubik's cubes, to different types of stress fields. Uniaxial tensile stress often leads to catastrophic lattice fracture (Figure 1a). For example, cracks in $La_2O_3$ ceramics initiate at 1.7% strain ($\varepsilon$) and rapidly propagate to complete fracture at $\varepsilon=1.8\%$, with virtually no plastic deformation[9]. While hydrostatic compression leads to uniform lattice shrinkage, collapse, or even phase transformation[10,11], as illustrated in Figure 1b. Both situations fail to provide the shear component necessary for dislocation nucleation and introduce no dislocation motion or slip of crystal planes. As a result, directly distorting the stiff lattice of HTS to generate dislocation networks remains a longstanding and formidable challenge.

Here, we propose an asymmetric stress field (ASF) strategy to overcome the intrinsic brittleness and structural rigidity of HTS. The ASF enables controlled lattice slip and twisting in stiff crystalline systems. As a proof of concept, we applied this strategy to $Ba_{1-x}K_xFe_2As_2$ (BaK122), a leading iron-based superconductor (IBS) for high-field applications[12], and successfully induced a high density of dislocations throughout BaK122 wires. These line defects serve as strong pinning centers, giving rise to a fivefold enhancement of $J_c$ at 33 T, a small anisotropy parameter, and a high irreversibility field, which are the remarkable features long sought after for advanced superconducting materials.

## 2. ASF-induced high-density dislocations

The generation of dislocations fundamentally requires a sufficient shear stress component acting along specific crystallographic planes to initiate atomic glide. However, traditional stress fields applied through processes such as drawing or rolling lack strong shear components. They typically result in dislocation densities that are several orders of magnitude lower than those found in metals and alloys. To address this challenge, we developed an innovative stress-modulation approach, allowing for the nucleation of a high-density of dislocations by tuning the stress components during deformation. As schematically illustrated in Figure 1c, we first apply sufficiently high hydrostatic pressure (red arrows) to suppress brittle fracture, and then deliberately break the stress symmetry along a specific direction to introduce asymmetric (non-hydrostatic) components (green arrows) and initiate directional shear. This concept is realized via a specialized extrusion technique that can be scaled up for mass production of superconducting long wires (Figure 1d and Figure S1a, fabrication details see Methods).

By carefully engineering the stress state of extrusion to combine fracture-suppressing pressure with lattice-shearing asymmetry, it becomes possible to overcome the limitations of brittle failure and activate controlled dislocation plasticity. Figure S1b quantifies the stress state generated by our ASF technique with a median deformation ratio. Finite element simulations demonstrate that our fabrication strategy not only achieves triaxial compressive characteristics but also exhibits high-stress and high-strain features. A macroscopic radial compressive strain of $\varepsilon_r$=65-75% was imposed on the BaK122 wire (not the crystal lattice) during the full extrusion process. The measured maximum compressive stress is ~230 MPa (Figure S1c), resulting in an enhancement of packing density from $\rho_0$=0.46 to the final $\rho_1$=0.97. It is remarkably different from the porous microstructures typically obtained through conventional processing routes with low stress (Figure S1c and d). As a result, grain connectivity is considerably improved, allowing for bulk supercurrents throughout the sample (Figure S1e).

Importantly, during the extrusion process, a key stress asymmetry develops near the outlet where axial confinement is rapidly released, while radial constraint is maintained by the sheath and die geometry. This imbalance disrupts the isostatic pressure state and introduces a significant shear component. The resulting asymmetric stress state generates off-diagonal elements in the local stress tensor, which activate lattice slip and twisting, as shown in Figure 1c. These shear-driven deformations enable the formation of dense dislocations without fracture. Consistent with this, distinct dimple-like features are observed on the fracture surface (Figure S1f), indicative of ductile behavior and substantial plastic deformation. This mechanism fundamentally differs from conventional symmetric processing routes, where the lack of directional shear severely limits dislocation nucleation in brittle, strongly bonded crystals.

The transmission electron microscopy (TEM) image in Figure 1e, taken along the normal direction (ND shown in Figure 3a) of the BaK122 wire, reveals a dense and highly entangled dislocation network in a single crystalline domain (Figure S2a). The disordered high-density dislocations reveal extensive lattice distortion, indicating that the applied asymmetric stress field effectively activates slip systems in the otherwise rigid crystal lattice. It underscores the ability of our strategy to overcome the intrinsic resistance to plastic deformation in brittle ionic/covalent systems. Figure 1f provides a side view of the dislocation lines (transverse direction, TD). The observed dislocations extend over several hundred nanometers, forming curved or intersecting configurations. Such long-range dislocation lines, along with the presence of pile-ups and junctions, suggest that the stress-driven deformation not only initiates dislocation nucleation but also facilitates their propagation and interaction, which are critical for the formation of a dislocation network.

Figure 1g compares the dislocation density, $\rho_D$, in our samples with that of metallic and non-metallic systems, including ceramics[13] and thermoelectric compounds[14,15]. Based on the statistical analysis of Figure 1e, the $\rho_D$ in our BaK122 wires is estimated to be ~$1.5 \times 10^9$ mm$^{-2}$, comparable to that achieved in irradiated samples[16,17]. It surpasses that of traditional non-metallic materials by two to three orders of magnitude and approaches levels commonly found in heavily deformed metals[18–20]. For comparison, BaK122 samples processed via conventional techniques show sparse dislocations (Figure S2b), with a density two orders smaller than our wires. Furthermore, we demonstrate the generality of this approach by applying it to a copper oxide superconductor, $(Cu,C)Ba_2Ca_3Cu_4O_{11+\delta}$ (Cu1234), which also exhibits a dense dislocation network with a density of ~$0.7 \times 10^9$ mm$^{-2}$ (Figure S2c). These results collectively highlight the capability of non-metallic materials to host ultra-dense dislocations when subjected to well-designed stress states.

Spherical aberration-corrected transmission electron microscope (TEM) was employed to examine the fine microstructural variations near dislocation lines. Interestingly, atomic displacements near the ASF-induced defects occur along a specific crystal plane, (113), where the Ba atoms exhibit a hexagonal close-packed configuration. The unaffected regions preserve the regular atomic ordering of Ba (Figure S3a). On the contrary, the slipped region exhibits evident atomic misalignments. As shown in Figure 1h, besides the unslipped region highlighted in blue, the Ba atoms on the upper right tend to move downward, while those on the left side slip to the lower left. The intensity profile of the atoms along the yellow line in Figure 1i suggests that the displacement of Ba atoms is approximately one angstrom. The measured twisting angle is about 0.5-1°. The corresponding schematic in Figure 1j and k illustrates the movement pattern of the pink Ba atoms at the top, implying the shear stress exerted on the crystal lattice. Similar slip patterns are also identified in Figure S3b and c, along with clear evidence of atomic glide viewed from alternative crystallographic directions (Figure S3d and e). All of these results indicate that the dislocation formation is primarily driven by the ASF-induced interlayer sliding and lattice twisting along the (113) slip system.

## 3. Dislocation dynamics in ASF and thermal activation

To further investigate the initiation and transformation mechanisms of dislocations induced by ASF, we analyzed the deformation process along with the subsequent annealing stages. First, we conducted *ab* initio molecular dynamics simulations (MDS) on the ASF, as presented in Figure 2a-c. As compressive strain increases, dislocations nucleate, propagate, and eventually entangle with each other. It leads to the formation of the high-density defects within the superconducting matrix at $\varepsilon=14\%$. In the meantime, amorphous regions emerge, potentially resulting in grain fracture. The calculated dislocation density increases to $\rho_D \sim 10^9$ mm$^{-2}$ when the strain reaches 15% (Figure 2f), consistent with Figure 1g. Electron-channeling contrast imaging (ECCI)[21] and TEM images further confirm the formation of high-density defects and the amorphization behavior during the ASF process (Figure S4 a-d). The simulation was terminated at $\varepsilon=16\%$ when lattice failure occurs, consistent with previous findings on CaFe$_2$As$_2$ micropillars exhibiting ~13.4% recoverable strain and fracture upon further deformation.[22] It suggests that the ThCr$_2$Si$_2$-type intermetallic compounds can tolerate substantial deformation before failure. Such large deformation tolerance is crucial for accommodating strain localization, which facilitates the formation of a high density of dislocations. Moreover, the strain limit is one order of magnitude greater than that of traditional ceramics (~1%), and comparable to that of the BN system that also benefits from twisted layers[23]. In the practical ASF process, local strain within crystalline grains during the ASF processing is estimated to reach ~20 %, causing significant grain refinements (Figure S4e-f).

Despite the high density, the entangled dislocation lines significantly scatter Cooper pairs and impair superconductivity. As a result, the complex defect structure introduced by ASF should be redesigned. In metals, recovery and polygonization are known to occur at temperatures of ~0.3$T_m$ ($T_m$ is the melting temperature), where thermally activated dislocation climb and rearrangement processes are initiated[7]. In contrast, non-metallic materials, such as ceramics, are generally considered insensitive to thermal activation due to their rigid bonding nature. However, we surprisingly observe that BaK122 exhibits significant defect evolution even under low thermal activation conditions. At temperatures below 400 °C (~0.28$T_m$, $T_m$~1400 °C for BaK122), dislocation structures undergo notable rearrangement, annihilation, and migration.

The dynamic behavior of dislocations under these conditions is confirmed by both MDS and in situ heating TEM observations. Figure 2d-e show the MDS of the microstructure evolution with the annealing time $t_0$, demonstrating the rearrangement and migration of dislocations. As shown in Figure 2f, the calculated dislocation density descends precipitously from the peak with annealing, reaching a plateau with $\rho_D$~5.5×10$^8$ mm$^{-2}$. This is further corroborated by the in situ heating TEM images in Figure 2h-l. The densely entangled dislocations in the as-deformed samples remain largely unchanged until the temperature reaches a threshold value, $T_1$=300 °C. Further heating above $T_1$ triggers the relaxation of dislocations via the annihilation of dislocations with opposite signs. Additionally, the dislocations collectively migrate downward by nearly 100 nm before being trapped by the grain boundary (GB)[24], which can also be found in Video 1. At the isothermal stage of 380°C, the development of dislocations at GBs is monitored over time. Figure S5 suggests that the prolonged annealing transforms the highly strained areas into alternating low-strain regions and defect regions, with a spacing of ~8 nm. Simultaneously, the zone axes of adjacent grains gradually align to form a low-angle grain boundary (LAGB). From a bulk perspective, the ASF-induced fine grains with submicron sizes undergo rapid growth driven by large grain boundary (GB) energy and higher temperature. This process involves active GB migration and reorientation, accompanied by the progressive formation of LAGBs during annealing, as supported by the calculated misorientation evolution shown in Figure 2g.

It should be noted that the in situ TEM characterization was conducted at relatively low temperatures to minimize potassium loss under high vacuum conditions. Nevertheless, BaK122 exhibits pronounced defect evolution. The intense dislocation dynamics resemble the recovery behavior commonly seen in metals[25], although the underlying mechanisms are influenced by the distinct crystal structures. In addition to the thermodynamic driving force provided by the high strain energy, the unique close-packed slip system is essential for the formation of a high density of dislocations and enabling their mobility. It suggests that IBS possesses sufficient lattice mobility under appropriate stress conditions to support thermally activated dislocation glide, further bridging the behavior between metallic and non-metallic systems.

**4. Strong vortex pinning landscape powered by ubiquitous dislocation arrays**

Building on the thermally driven dislocation dynamics observed in the in situ TEM studies, we applied a straightforward annealing treatment to further regulate the dislocation structures both within grains and along grain boundaries. As shown in Figure 3b, the BaK122 grains are closely connected by clean GBs. The highly compacted submicron grains promote grain coarsening and facilitate thorough element diffusion between grains during annealing, which may prohibit the nucleation of the GB wetting phase[26]. More importantly, all the GBs are decorated with periodic dislocations. Figure 3c-d illustrate the dislocation array at the GB in region A, viewed from the [001] direction. The dislocation cores have a diameter of 3 nm with an interspacing of 10 nm ($\rho_D$ ~$5\times10^8$ mm$^{-2}$). The Geometric Phase Analysis (GPA) in Figure 3e proves a complex strain state within and around the dislocation cores, leading to the distorted lattice fringes in the surrounding regions. In IBS, the superconducting transition temperature ($T_c$) is highly sensitive to the crystal lattice, such as the height of As atoms from the Fe plane and the As-Fe-As bond angle[27]. Excessive local stress or strain may suppress $T_c$ or even lead to the formation of non-superconducting regions, which act as vortex pinning centers. Moreover, the selected area electron diffraction (SAED) in the inset of Figure 3c indicates a misorientation angle of $\theta_m$~3.4°, much smaller than the critical value of $\theta_c$~9° for undisturbed intergrain currents[28]. Step-like dislocation arrays are also found at LAGBs in Regions B and C (Figure S6). Electron energy loss spectroscopy (EELS) mapping (Figure S6h-k) reveals no significant element segregation near the dislocations, ruling out compositional variations as a cause of the lattice distortion. High-density dislocation arrays can also be seen at GBs in the TEM images viewed from the TD (Figure S7). According to the thermal dynamic analysis in Figure 2, it is concluded that the dislocations migrate to GBs and self-organize into defect arrays. The size of the strained areas at GBs is close to the $\xi$ of BaK122 ($\xi$~3.4 nm)[29], making them strong pinning centers. Meanwhile, the less-strained regions between dislocations retain superconductivity and serve as effective pathways for supercurrent transport, as shown in Figure 3f. This spatial coexistence of distorted and undistorted regions highlights the functional architecture of the dislocation arrays, which balance strong vortex pinning with continuous current flow.

The moderate thermal activation has not eliminated all the dislocations within grains. As observed in Figure 3g-k and Figure S7a-b, plenty of dislocation lines are visible in the TEM images taken along ND and TD. The short segments, about 4 nm wide and 30 nm long, observed in Figure 3h are actually the endpoints of dislocation lines, with the viewing direction nearly parallel to the line defects. The GPA in Figure 3i demonstrates the strained lattice near the dislocations. When viewed along the TD, dislocations with different lengths are visible, as shown in Figure 3j. Some dislocation lines align almost parallel to the ND, while some form an angle of 20~40° to the ND (Figure S7). This is consistent with the inclined twisting plane in Figure 1k.

Based on the **g·b**=0 invisibility criterion, the dislocation contrast disappears under **g** = [0$\bar{1}$3] (Figure S8b). As a consequence of the fact that the dislocation lines are initiated within the (113) slip plane, we determine the Burgers vectors to be **b** = [6$\bar{3}\bar{1}$]. As illustrated in Figure S8c, the dislocations are color-coded by type, revealing that most of them are of the mixed character, with a small portion showing screw-like features. This is similar to that before annealing, except that the dislocation lines are more entangled, the slip directions are more disordered, and the slip bands are wider. These features make it difficult to determine the exact dislocation types, though they are overall of mixed character. After annealing, the slip becomes more coherent and the slip band width is reduced.

As marked in Figure 3k, the width of the dislocation lines observed from TD is 2.5-4.3 nm. The visible portions of some dislocation lines extend over several hundred nanometers (Figure 3j and Figure S7a), enabling better pinning for the longer sections of vortices. The average distance between dislocations is ~60 nm. This is in agreement with the estimated intragranular dislocation density ($\rho_D$~1.2×10$^8$ mm$^{-2}$) derived from the side-viewed TEM images. The defect density is 2-4 orders of magnitude higher than that of the ceramics prepared with traditional processes (~10$^4$-10$^6$ mm$^{-2}$)[13]. These high-density tilted line defects, with transverse dimensions close to $\xi$, create localized distortions and strain fields in the crystal lattice. They break the long-range coherence of the superconducting state and act as strong pinning centers. The ASF approach generates pervasive line defects throughout the entire sample, triggering a profound transformation in the vortex-pinning landscape.

Figure 4 compares two disparate pinning scenarios of HTS wires. For the common wires without pinning centers at GBs (Figure 4a and b), the intragrain pinning is dominated by sparse point defects possibly caused by atomic vacancies[30,31]. According to the model proposed by Gurevich and Cooley, vortices preferentially enter GBs and transform into the Abrikosov-Josephson (AJ) vortices[32]. AJ vortices are weakly pinned by GBs and easily flow along the percolation channel. Consequently, the transport $J_c$ at low fields is severely compromised, as reflected by the field-up branch of the $J_c$-$B$ curves at low fields (Figure 4e). As the external field increases, vortices enter the grains and form Abrikosov lattices which are pinned by intragrain defects. The Abrikosov (A) vortices magnetically interact with the AJ vortices at GBs and prevent their dissipative motion. Consequently, the transport $J_c$ is improved as the field increases. After all the AJ vortices are captured by the A vortices, they are collectively pinned and moved under ideal conditions. As a result, the deviation at $B_d$ between the field up and field down branches of the $J_c$-$B$ curves reflects the GB pinning strength in polycrystalline wires. In the ASF-processed wires, the dissipation at GBs is largely ameliorated after dislocation pinning arrays are introduced (Figure 4c and d). The motion of the AJ vortices is mostly prevented both at low and high fields. Therefore, the $B_d$ of the ASF wires is substantially depressed when compared to the Bi2223[33] and hot-pressed (HP) BaK122 wires[34].

For the BaK122 wires fabricated by traditional processes[35], the $J_c$ anisotropy depends on the intrinsic anisotropy of electron mass because of the dominant random point defects, as depicted in the right inset of Figure 4f. These low-density point defects provide non-directional weak collective pinning, leading to limited $J_c$ and strong reliance on intrinsic anisotropy. The intrinsic anisotropy parameter of BaK122 can be expressed as $\gamma=(m_c/m_{ab})^{1/2}=H_{c2}^{ab}/H_{c2}^{c}$, where $H_{c2}$ is the upper critical field, $m_c$ and $m_{ab}$ are respectively the effective masses of electrons along the $c$-axis and the $ab$-plane[36]. This value is generally $1<\gamma<2$, indicating that the $H_{c2}$ and the $J_c$ are larger when $B//ab$ ($\theta=90°$, strong $c$-axis texture of the grains). It is consistent with the measured anisotropy behavior of the traditional BaK122 wire shown in Figure 4f. On the contrary, there is a reversed anisotropy in the ASF wires, $J_c(90°)<J_c(0°)$, suggesting the preponderant contribution from the *quasi c*-axis correlated defects observed in the TEM images, as shown in the left inset of Figure 4f. Moreover, the tilt dislocation lines also contribute to the vortex pinning and the $J_c$ enhancement with $\theta=90°$. The $J_c(\theta=90°)$ achieves $2.2\times10^5$ A/cm² at 4.2 K and 10 T, much larger than the reported record value[37]. Consequently, the $J_c$ anisotropy remains below 2 even with the directional pinning centers. The small anisotropy facilitates the design of homogeneous magnets, of which the performance is greatly limited by the lower $J_c$ of the anisotropic conductor[38].

**5. Ultrahigh critical current density at high magnetic fields**

Based on the TEM and MDS results, it can be concluded that the pinning structure of the ASF wires is determined by the thermal-activated transformation of dislocations. To investigate the influence on the final performance, the wires were annealed at 880 °C for 20 minutes (S20), 60 minutes (S60), and 300 minutes (S300). The kernel average misorientation (KAM) maps in Figure S9, which are commonly applied to evaluate crystal defects[39], unveil a high density of dislocations within grains and at GBs. As annealing continues, the intragrain defects migrate to GBs. The fully annealed S300 sample even shows few defects in grains. The inset of Figure 5a manifests an exponential decay of the average *KAM* values with annealing time, indicating a rapid reduction in defect density. The main panel of Figure 5a describes the field-dependent $J_c$ of the three samples. Only the *B*//TD configuration is discussed here because the $J_c(B$//ND) exceeds the maximum output of our current source at low fields. S20 shows a lower self-field $J_c$ and a faster drop with fields than the other two samples, which can be ascribed to the weak grain coupling caused by insufficient annealing. The $J_c(0\ T)$ of S300 achieves the same level as that of S60 but exhibits a faster decline in $J_c$ under field due to the reduced intragrain defects. In contrast, S60 exhibits a field-independent $J_c$ up to an accommodation field $B^*=3.2$ T, followed by a power law decay at high fields. This is an archetypal phenomenon for strong vortex pinning, which has not been observed in previously reported IBS wires. It means that the constructed pinning landscape of tilt dislocations provides sufficient strong pinning sites to efficiently accommodate magnetic vortices up to 3.2 T, even when the field is parallel to TD. The calculated pinning center density from $B^*$ is approximately $1.5\times10^9$ mm$^{-2}$, close to the TEM and MDS results. Therefore, the robust field-dependent $J_c$ of S60 is attributed to the strongly coupled grains reinforced by well-established high-density inclined line defects.

With the optimized pinning structure, the ASF wire presents an extraordinarily large irreversibility field ($B_{irr}$), at which the dissipationless current vanishes. As shown in Figure 5b, the $B_{irr}^{//c}$ of the ASF wires exhibits a precipitous ascent with cooling, reaching 46 T at 26 K. A rough estimate of $B_{irr}^{//c}(T)$ using a power law dependence suggests a high value of 120 T at liquid helium temperature, which explains the robust field dependence of $J_c$. More importantly, the $B_{irr}$ of the ASF wires not only overwhelms that of NbTi, Nb$_3$Sn, and MgB$_2$ but also outperforms Bi-based and NdFeAsO$_{1-x}$F$_x$ superconductors below 35 K[40], indicating competitive advantages in high-field applications. The $B_{irr}$-$t$ curve ($t$=T/T$_c$) in Figure S10d unravels the thermodynamic behavior of vortices among different HTS. The ASF wires present a sharper increase of $B_{irr}$ with cooling than other HTS. This is because the thermal fluctuation quantified by the Ginzburg number (Gi) is less pronounced in BaK122 (Gi~$10^{-4}$) than that of NdFeAsO$_{1-x}$F$_x$ (Gi~$10^{-3}$-$10^{-2}$) and cuprates (Gi~$10^{-2}$), resulting in a slower vortex creep[41]. As for the strongly anisotropic Bi-based superconductors, the 3D vortex lines transition to easily movable 2D pancake vortices, leading to a considerably depressed $B_{irr}$ at intermediate and high temperatures[42].

Benefiting from the strong pinning landscape and the large $B_{irr}$, the ASF wires present outstanding current-carrying capability. First, they exhibit sharp superconducting transitions in the I-V curves ($n$-values>60), indicating a highly homogeneous superconductor with strong vortex pinning (Figure S10a and b). A critical current ($I_c$) of 325 A is achieved at 4.2 K and 10 T in our 4 mm wide sample. As shown in Figure 5c, the $J_c$(4.2 K, 10 T) is 4.5×$10^5$ A/cm$^2$, which is the highest value ever reported for IBS wires. The dense dislocations exhibit a prominent pinning efficiency at high fields. Therefore, the $J_c$ presents a robust field dependence and remains 2×$10^5$ A/cm$^2$ at 33 T, which is five times that of the former record of the HP BaK122 wire[37]. The $J_c$ of the ASF wires exhibits a prominent advantage over NbTi, Nb$_3$Sn, and MgB$_2$ wires above 20 T, which demonstrate a relatively poor $J_c$-$B$ characteristic, with their $J_c$ rapidly dropping below the threshold of $10^5$ A/cm$^2$ for practical applications. Moreover, the ASF wires outperform Bi2223 in the whole testing range[43] and even surpass that of as-grown BaK122 single crystals dominated by weak collective pinning[17]. The vortex pinning force density, $F_p=J_c\times B$, of the ASF wire achieves 65 GN/m$^3$ at 25 T. The $F_p$ of the ASF wires at fields over 10 T is much higher than other superconductors shown in Figure S10f. Additionally, correlated defects enhance pinning energy by interacting with more segments of vortex lines[40], thereby providing better resistance to thermal fluctuations at elevated temperatures. As shown in Figure 5d and Figure S10e, the $J_c$(10 T) of the ASF wire still achieves $3\times10^5$ A/cm$^2$ and $1.1\times10^5$ A/cm$^2$ at 10 K and 20 K, respectively. It far exceeds that of Bi2223 and MgB$_2$ wires at 20 K, laying the foundation for the potential use of IBS as a practical superconductor for high-field applications at intermediate temperatures accessible with cryocoolers.

Figure 5e summarizes the $J_c$ development of IBS wires over the last decade[44]. Although the $J_c$ performance was rapidly enhanced by synergistically improving the density and texture of the superconducting core, it has reached the bottleneck (blue dashed line, $J_c\sim1.5\times10^5$ A/cm$^2$) since 2018 due to the impaired grain connectivity by GB impurities[45]. The ASF strategy simultaneously overcomes the long-standing issue of weak GB coupling and the critical challenge of insufficient strong pinning centers in IBS wires, resulting in a three-fold enhancement in $J_c$ at 10 T[37]. In contrast, directly doping nanoparticles into PIT-processed IBS wires is challenging due to the solid-state reaction process and uncontrolled grain growth during high-temperature sintering. Furthermore, our approach relies on the simple and scalable powder-in-tube method, which is generally considered cost-effective for fabricating iron-based superconducting wires[46]. Multiple outer sheaths with low cost and high mechanical strength have already been bonded during fabrication, providing robust mechanical support to withstand huge Lorentz forces under high fields. All of these advantages enable the stable operation of high-performance IBS wires in high magnetic fields.

**6. Conclusion**

In conclusion, we propose an asymmetric stress field strategy to overcome the inherent brittleness and lattice rigidity of iron-based superconductors. By superimposing shear stress onto hydrostatic compression, this approach activates controlled lattice slip and twisting, enabling the generation of a high density of dislocations. In situ transmission electron microscopy studies reveal a metal-like recovery behavior of dislocations in this non-metallic system during the subsequent thermally activated process. It enables further tailoring of dislocations into effective and uniformly distributed strong pinning centers of magnetic vortices. Benefiting from such a strong pinning landscape of tilted dislocation lines, the IBS wires exhibit ultrahigh $J_c$, large $B_{irr}$, and low anisotropy, making them exceptional candidates for commercial fusion reactors, next-generation accelerators, and high-field MRI systems. Our counterintuitive strategy not only advances vortex pinning engineering in HTS, but also offers a generalizable route for directly manipulating dislocation structures in rigid crystalline systems across a wide range of functional materials.

## 7. Experimental Section/Methods

**Sample Preparation:**

The BaK122 precursor bulks with the nominal composition $Ba_{0.6}K_{0.5}Fe_2As_2$ were prepared by the solid-state-reaction method. The ASF wires were fabricated using the powder-in-tube (PIT) method. Precursor powders were packed into Ag tubes (outer diameter: 8 mm; inner diameter: 4.0 mm). The filled billets were extruded in a single pass to produce wires with an outer diameter of 2.0-2.4 mm, then drawn into 0.98 mm wires. These were subsequently flat-rolled into 0.4 mm-thick tapes and inserted into flat-rolled stainless steel (SS) tubes. The resulting composites were further rolled into tapes with a final thickness of 0.8 mm. This ASF-based process achieves a 93% cross-sectional area reduction in one extrusion step, corresponding to an extrusion ratio of ~16 (defined as the ratio of the initial cross-sectional area $S_0$ to the final area $S_1$, *i.e.*, $R = S_0/S_1$). In contrast, conventional IBS wire processing typically achieves only ~10% area reduction per drawing pass, requiring over twenty passes to reach a 0.98 mm diameter. Our extrusion-based method thus significantly streamlines the fabrication process.

All tapes were sintered in an Ar atmosphere at 880 °C for one hour, unless otherwise noted. For comparison (*e.g.*, Figure 5a), samples were also sintered at 880 °C but with different dwell times (20 min, 60 min, 300 min) to study the annealing effect.

**Transport Critical Current Measurements:**

Transport critical current ($I_c$) was measured using the standard four-probe method with a criterion of 1 µV/cm. The critical current density ($J_c$) below 14 T was obtained using a superconducting magnet at the Institute of Plasma Physics, Chinese Academy of Sciences (CAS). For fields under 10 T, $J_c$(4.2 K) was measured with the magnetic field parallel to the tape surface (B//TD) due to current source limitations. Angle-dependent $J_c$ data between 14 T and 25 T were measured using a hybrid magnet at the High Field Laboratory for Superconducting Materials, Tohoku University. Data above 26 T were obtained using a water-cooled magnet at the High Magnetic Field Laboratory, Chinese Academy of Sciences.

**Microstructural Characterization:**

Electron channeling contrast imaging (ECCI) was performed using a Zeiss MERLIN field emission scanning electron microscope. Grain size and texture of the superconducting cores were characterized via electron backscatter diffraction (EBSD) equipped on a scanning electron microscope (ASFM, Zeiss SIGMA). The sub-grain structure was assessed by EBSD-based Rotation Angle maps (misorientation<5º), and Kernel Average Misorientation (KAM) maps were used to qualitatively evaluate local defect densities.

Transmission electron microscope (TEM) images were acquired using an aberration-corrected high-resolution electron microscopy (JEMARM200F, JEOL) operated at 200 kV with an imaging aberration corrector (CEOS). Samples were prepared using a focused ion beam (FIB) system. Elemental uniformity was analyzed via electron energy loss spectroscopy (EELS) and energy-dispersive spectroscopy (EDS). Strain mapping was performed using geometric phase analysis (GPA) in Gatan Digital Micrograph. Multiple high-resolution TEM images were collected from at least three regions per sample for statistical analysis of dislocation density.

For in situ heating experiments, FIB-prepared samples were transferred to microelectromechanical-systems-based heating chips (Nano-Chip, DENSsolutions) on a DENSolution DH30 holder. Heating steps included 300 °C, 360 °C, and 380 °C with dwell times of 10 min, 20 min, and 40 min at each stage, respectively.

**Physical Property Measurements:**

Magnetic properties of the bare superconducting cores were measured using a magnetic property measurement system (MPMS). Field-dependent resistance was measured up to 55 T in a pulsed magnet at Wuhan National High Magnetic Field Center. The irreversibility field ($B_{irr}$) was determined at the criterion of 10% of the normal-state resistance. Magneto-optical (MO) imaging was conducted using a Montana Instruments optical cryostat with a custom-built magnet providing perpendicular fields.

**Molecular Dynamics Simulations:**

Ab initio molecular dynamics simulations of dislocation behavior in BaK122 were performed using the LAMMPS package[49]. The interatomic potential was based on a hybrid Morse and modified embedded atom method (MEAM) model. Simulation cells were 50×50×50 nm$^3$ containing ~1.25 million atoms, with a time step of 0.1 fs. Simulations were carried out at 300 °C to match the experimental deformation condition. A compression strain of 0.20 was imposed as the target external condition. Additionally, annealing effects at 880 °C were simulated on structures pre-strained to $\varepsilon$=0.14. Dislocation multiplication and GB misorientations were analyzed. All visualizations were generated using OVITO[50,51].

**Acknowledgements**

This work is partially supported by the National Key R&D Program of China (Grant Nos. 2018YFA0704200 and 2017YFE0129500), the National Natural Science Foundation of China (Grant Nos. 51861135311, U1832213, 51977204, 51721005, 52172275, 52107031, and 51802189), the Strategic Priority Research Program of Chinese Academy of Sciences (Grant No. XDB25000000), the International Partnership Program of Chinese Academy of Sciences (Grant No. 116GJHZ2023005MI), Key Research Program of Frontier Sciences of Chinese Academy of Sciences (QYZDJ-SSW-JSC026), Beijing Municipal Natural Science Foundation (Grant No. 3222061), Natural Science Foundation of Shandong Province (Grant No. ZR2021ME061).

The authors would like to thank Dr. Peng Xu, Prof. Huajun Liu, Prof. Jinggang Qin, Dr. Donghui Jiang, and Prof. Chuanying Xi at Hefei for $I_c$–$B$ measurement. The authors would like to thank Prof. Kai Wang, Dr. Chen Guo, Dr. Peng Yang, Dr. Cong Liu, Dr. Xudong Zhang, Dr. Caida Fu, and Dr. Minghui Tang of the Institute of Electrical Engineering, Chinese Academy of Sciences for analyzing and discussing results. The authors wish to acknowledge Prof. Jihua Huang, Dr. Zheng Ye, Dr. Zhenqian Lang, and Ms. Ruobing Zhao of the University of Science and Technology Beijing for their help with the sample fabrication.

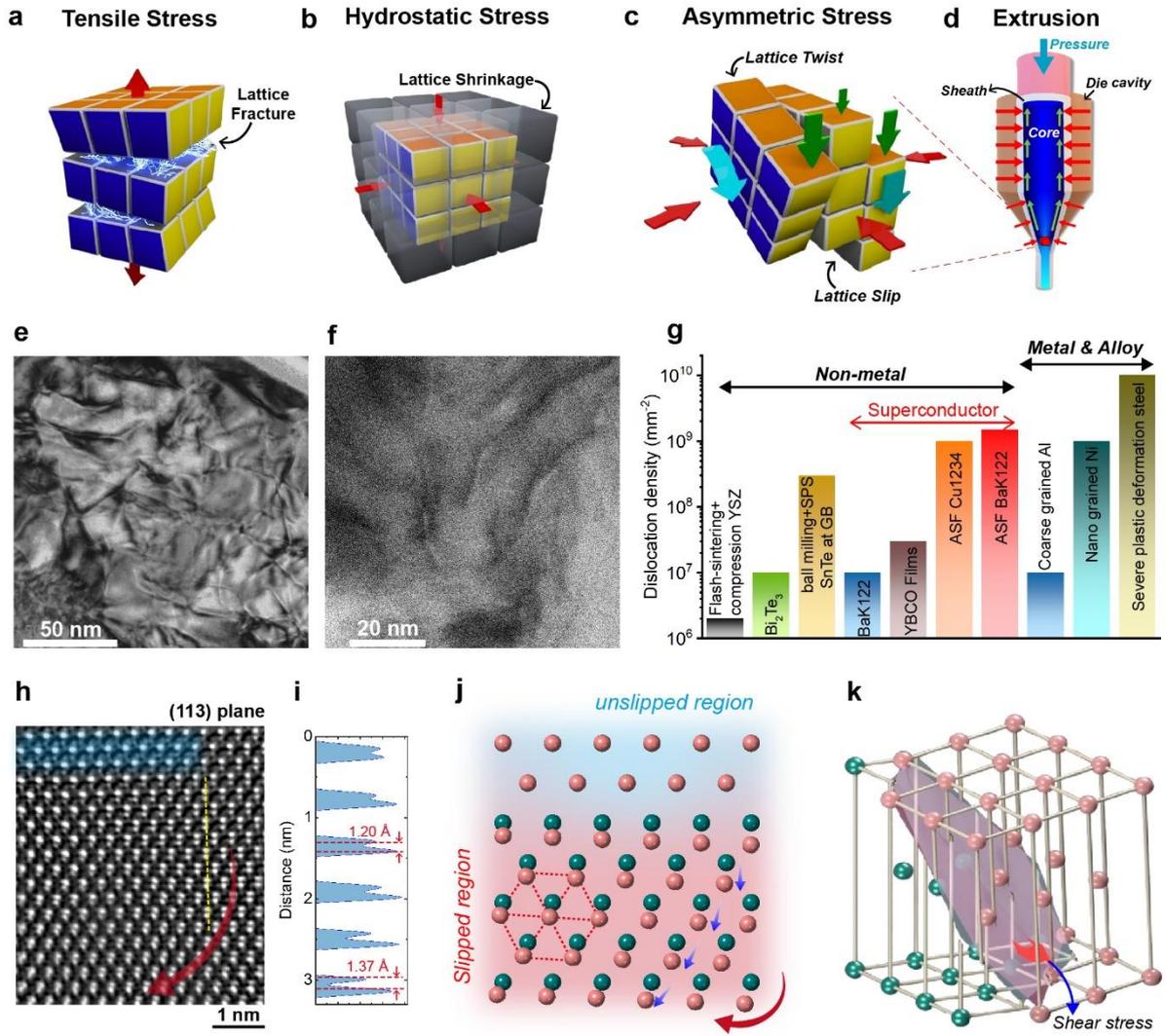

**Figure 1. Generation of a high density of dislocations in rigid crystal lattices by asymmetric stress field. a-c**, Schematic illustration of the lattice response to different stress states, represented using a Rubik's cube model. **a**, Uniaxial tensile stress leads to brittle fracture. **b**, Hydrostatic compressive stress results in uniform lattice shrinkage. **c**, The proposed asymmetric stress field (ASF), combining hydrostatic pressure (red arrows) and directional shear (green arrows), induces lattice slip and interlayer twisting without fracture. **d**, schematic of the extrusion process that realizes the ASF strategy. During deformation, radial hydrostatic compression is applied through the die cavity (red arrows), while longitudinal stress relaxation induces shear components along the wire axis (green arrows), establishing an asymmetric stress state. Notably, the shear component intensifies near the die exit due to the abrupt release of axial constraints, facilitating lattice slip. TEM image viewed along the **e**, normal direction (ND) and **f**, transverse direction (TD). **g**, Dislocation density ($\rho_D$) of BaK122 processed by the ASF strategy, compared to various non-metallic and metallic systems. **h**, HAADF-STEM image of the (113) slip plane showing atomic displacements in BaK122. **i**, Intensity profiles along the

yellow dashed line reveal Ba atom shifts of ~1 Å. **j**, Schematic diagram showing lattice slip along the (113) plane. Blue and red regions denote the unslipped and slipped areas, respectively. **k**, Atomic model illustrating the activation of shear-driven slip and twisting along the (113) plane under the applied asymmetric stress field.

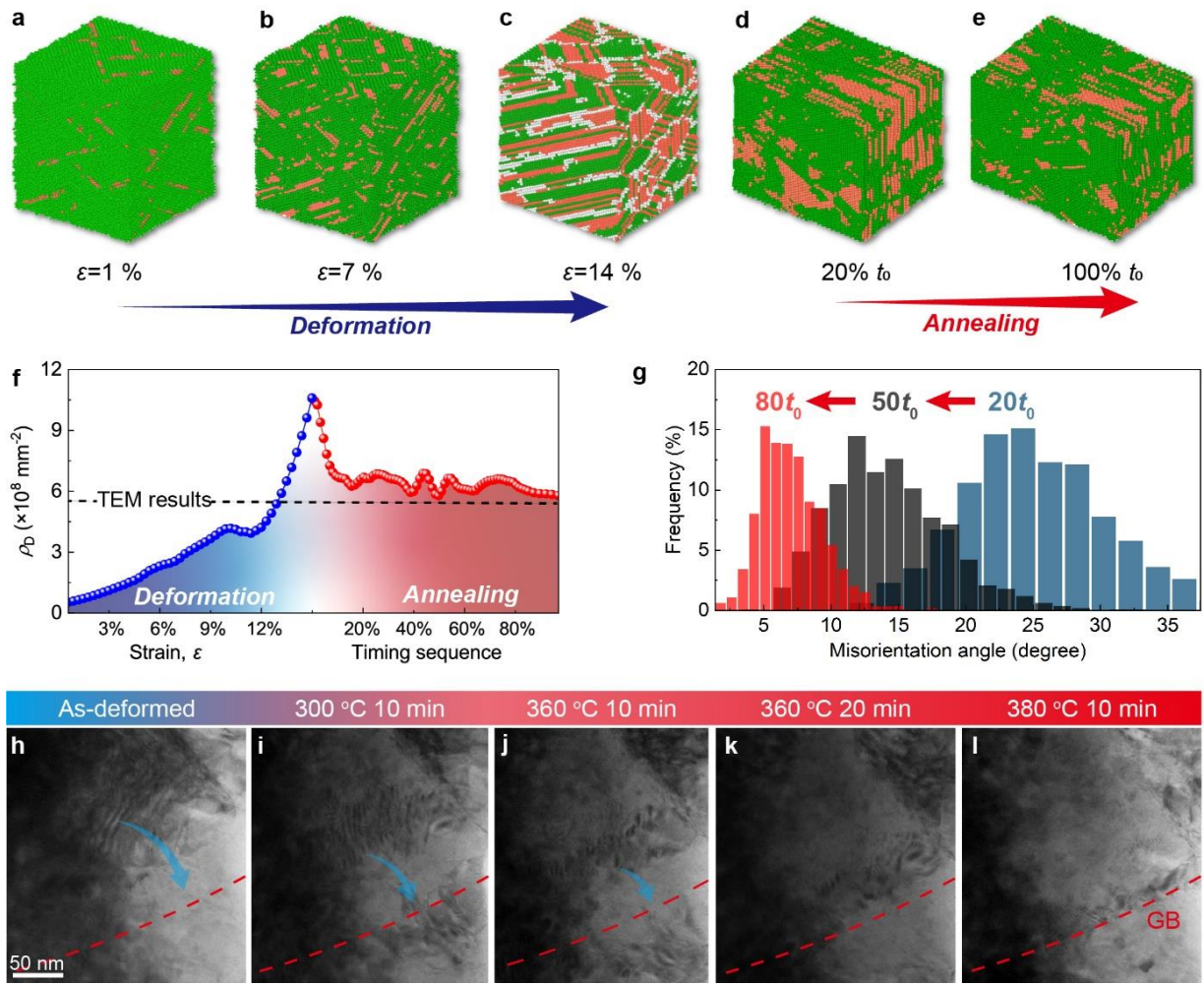

**Figure 2. Dislocation dynamics during ASF and thermal activation.** The *ab* initio molecular dynamics simulations on the evolution of dislocations during (**a-c**) the extrusion and (**d-e**) annealing processes ($\varepsilon$ is the strain, $t_0$ is the annealing time). The top surface is perpendicular to the [001] direction of the BaK122 lattice. The green region represents the initial undisturbed structure, and the red is the dislocation region. The white denotes the amorphization area. **f**, Dislocation density as a function of strain and annealing time. **g,** Distributions of GB misorientation angle at different annealing stages. **h-l**, in situ heating TEM images exhibiting the transformation of dislocations. The blue arrow indicates the moving direction of dislocations, and the red dashed line marks the GB.

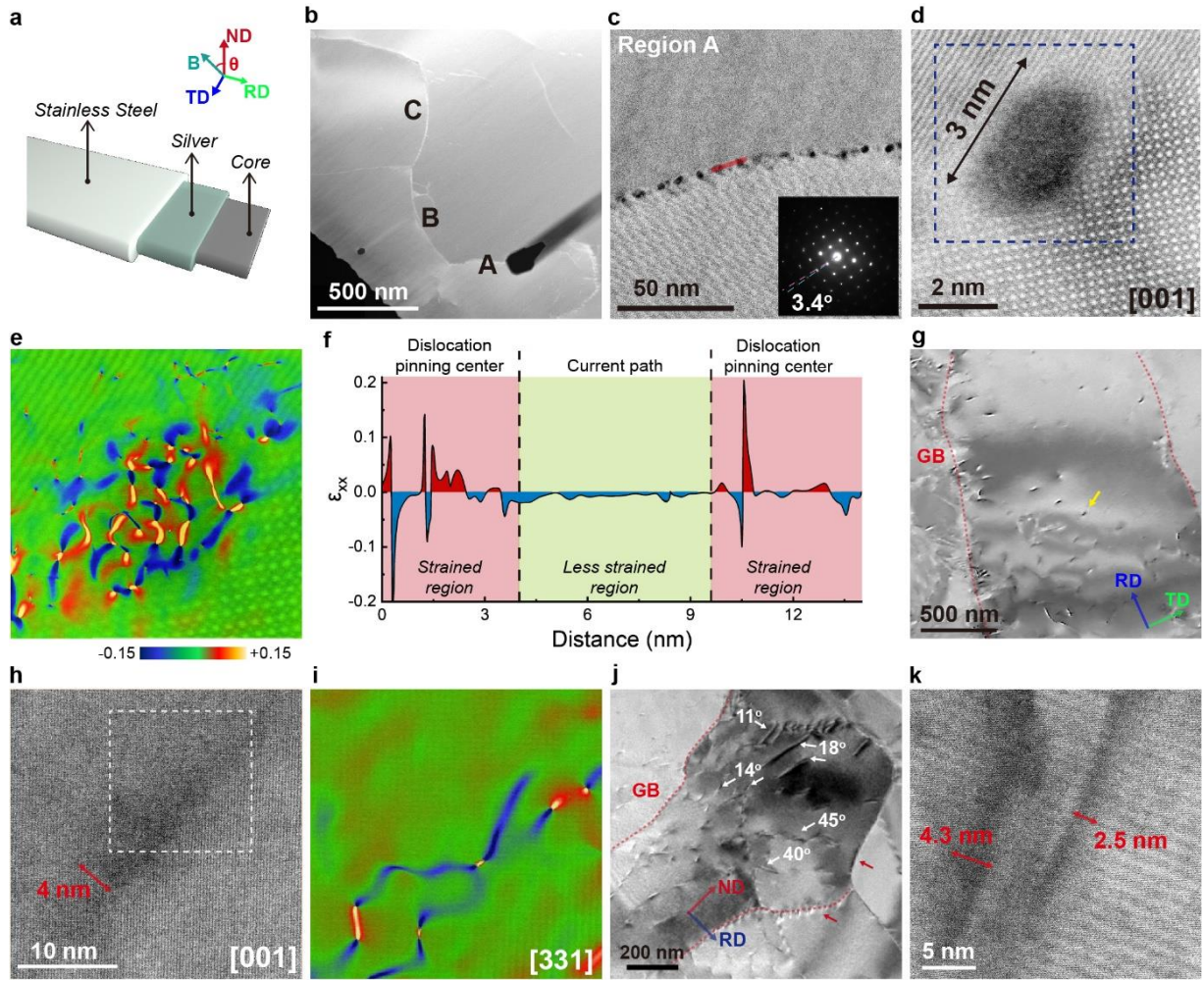

**Figure 3. Microstructure of dislocations in grains and at GBs. a**, Schematic diagram of the stress-engineered stainless steel/Ag composite wires. The directions of the wires are shown on the top right. The angle between ND and the external field B is denoted as *θ*. **b**, HAADF-TEM images of the BaK122 superconducting core viewed along the [001] zone axis. **c**, Dislocations at GB in region A. The inset is the corresponding SAED image. **d**, Enlarged view of the dislocation cores in (**c**). **e**, Strain maps of the dashed box in (**d**). **f**, Strain profile along the red line in (**c**). Intragrain dislocation lines observed from (**g**) the ND and (**j**) the TD. (**h**) and (**k**) are the enlarged views of the dislocations viewed from ND and TD, respectively. **i**, Strain maps of the dashed box in (**h**), with the same color scale as in (**e**).

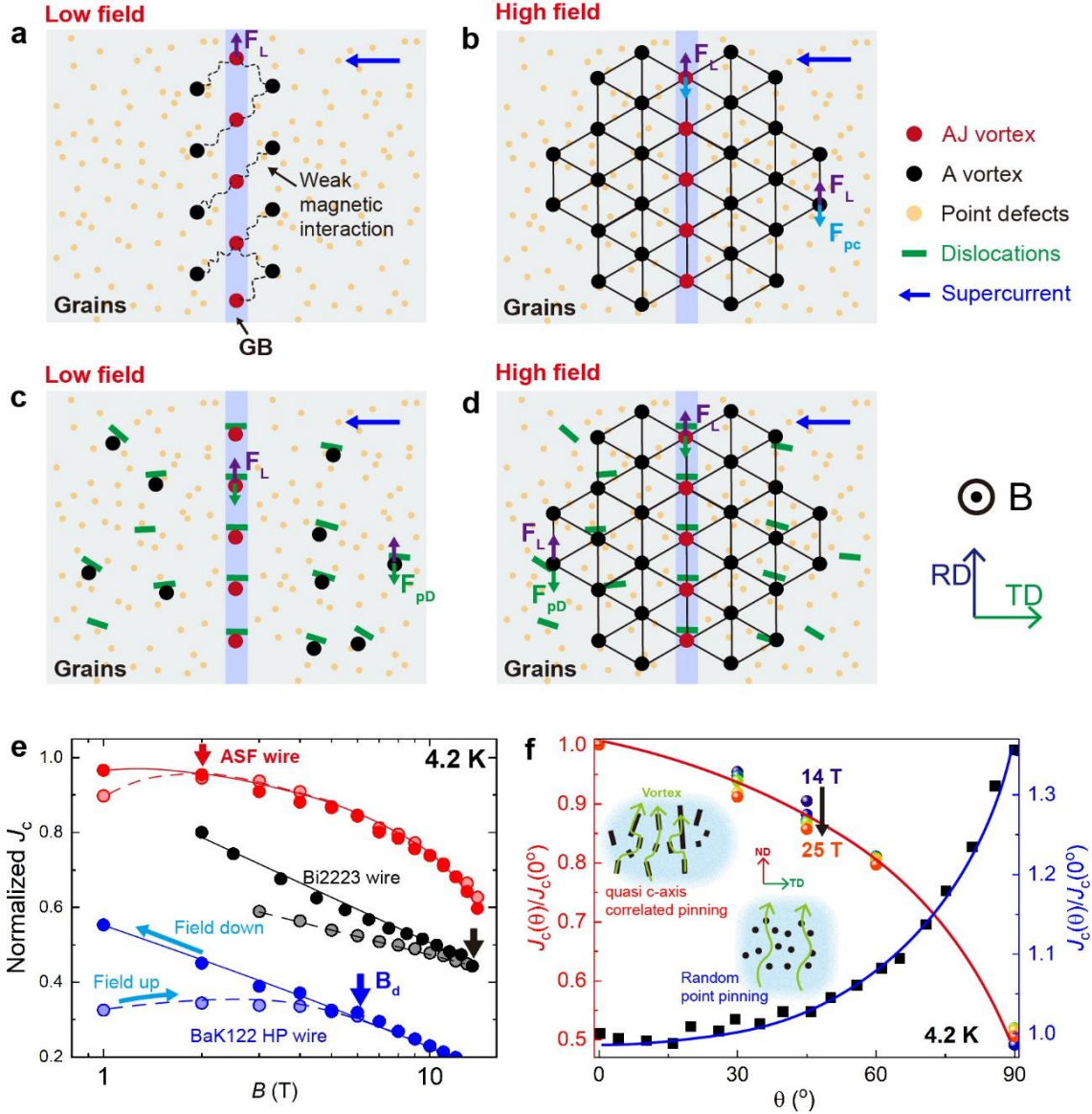

**Figure 4. Vortex pinning mechanism of the ASF-processed IBS wires. a-d**, Schematic diagram explaining the vortex pinning mechanism at low fields (a and c) and high fields (b and d). Vortex pinning structures (a-b) without and (c-d) with dislocations. $F_{pc}$, collective pinning force by point defects, $F_{pD}$, strong pinning by dislocations. **e**, $J_c$(4.2 K)-$B$ hysteresis plotted with a logarithmic $B$-axis. $J_c$ values are normalized to the zero-field value in the decreasing field branch for the ASF and HP wires, and to the $J_c$(2 T) for the Bi2223 wire. The Bi2223 and HP data are vertically shifted for clarity. The arrows mark the $B_d$ where the field-down branch (solid lines) begins to deviate from the field-up branch (dashed lines). **f**, Angular dependence of $J_c(\theta)/J_c(0°)$ for the ASF wire (left axis, red line) and the traditional IBS wire (right axis, blue line, $B$=0.5 T)[35], $\theta$ is marked in Figure 3a. The left and right insets depict the vortex pinning scenarios for the *quasi c*-axis correlated defects and random point defects, respectively.

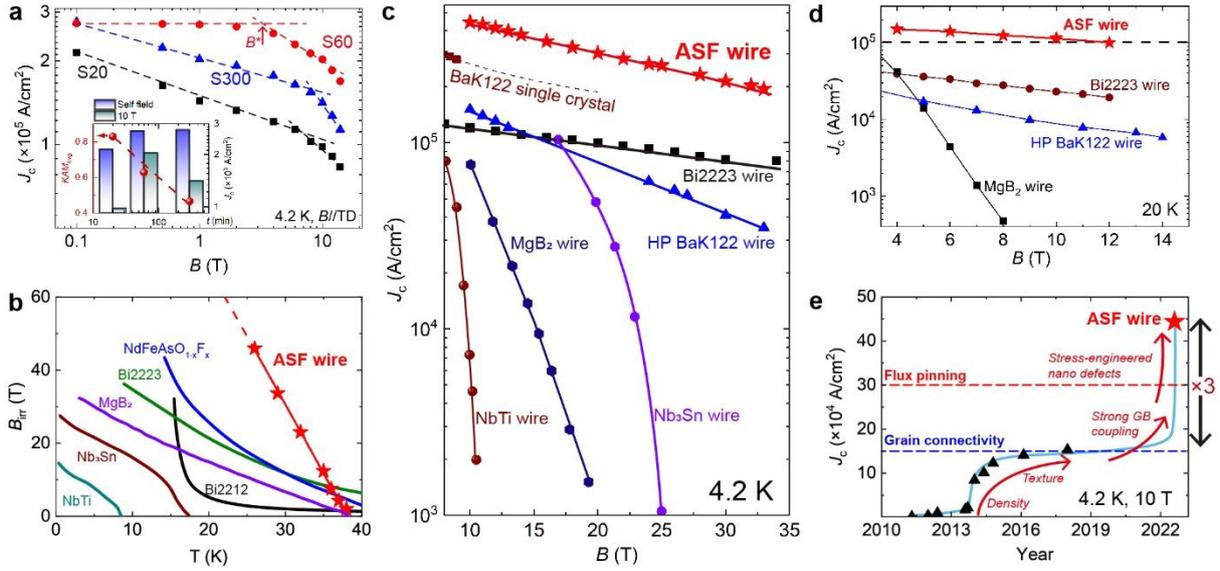

**Figure 5. Exceptional superconducting properties of the ASF-processed IBS wires. a**, Field dependence of $J_c$($B$//TD, 4.2 K) on double-logarithmic scales for samples S20, S60 and S300. The red arrow marks the accommodation field $B^*$ below which the $J_c$ is almost field-independent. Inset: Average Kernel Average Misorientation ($KAM_{avg}$, left axis) as a qualitative indicator of defect density, and $J_c$ (right axis) at self-field and 10 T as a function of the annealing time. **b**, Temperature-dependent irreversibility field, $B_{irr}^{//c}$. The dashed curve in (**b**) is the fitting result using $H_{irr}(T)=H_{irr}(0)(1-T/T_c)^m$, where m is the power law exponent. Field dependence of $J_c$($B$//ND) at (**c**) 4.2 K and (**d**) 20 K [34,47,48]. **e**, Development of $J_c$(4.2 K, 10 T) of IBS wires. The blue dashed line denotes the $J_c=1.5\times10^5$ A/cm$^2$, which is limited by poor grain connectivity. The red dashed line ($J_c=3\times10^5$ A/cm$^2$ for Ba$_{0.6}$K$_{0.4}$Fe$_2$As$_2$ single crystals) represents the limitation due to weak vortex pinning.

27